\documentclass[a4paper,fleqn,usenatbib]{mnras}

\usepackage[T1]{fontenc}
\usepackage{ae,aecompl}

\usepackage{graphicx}	
\usepackage{amsmath}	
\usepackage{amssymb}	

\title[Spectroscopic survey of emission-line stars. {I. B[e] stars}]{Spectroscopic survey of emission-line stars. I. B[e] stars\thanks{Based on observations 
collected with the Perek 2-m telescope at Ond\v{r}ejov Observatory, Czech Republic}}
\author[A. Aret et al.]{
A.~Aret$^{1}$\thanks{E-mail: anna.aret@to.ee (AA); michaela.kraus@asu.cas.cz (MK); miroslav.slechta@asu.cas.cz (MS)},
M.~Kraus,$^{1,2}$\footnotemark[2] 
and M.~\v{S}lechta$^{2}$\footnotemark[2]
\\
$^{1}$ Tartu Observatory, 61602, T\~oravere, Tartumaa, Estonia\\
$^{2}$ Astronomick\'y \'ustav, Akademie v\v{e}d \v{C}esk\'e republiky,
Fri\v{c}ova 298, 251\,65 Ond\v{r}ejov, Czech Republic
}

\date{Accepted 2015 November 20. Received 2015 November 20; in original form 2015 August 11}

\pubyear{2015}

\begin{document}
\label{firstpage}
\pagerange{\pageref{firstpage}--\pageref{lastpage}}
\maketitle

\begin{abstract}
Emission-line stars are typically surrounded by dense 
circumstellar material, often in form of rings or disc-like structures. Line emission 
from forbidden transitions trace a diversity of density
and temperature regimes. Of particular interest are the forbidden lines of 
[O\,{\sc i}] $\lambda\lambda$6300, 6364 and [Ca\,{\sc ii}] $\lambda\lambda$7291, 7324. 
They arise in complementary, high-density environments, such as the 
inner-disc regions around B[e] supergiants. To study physical conditions traced by these
lines and to investigate how common they are, we initiated a survey of emission-line stars. 
Here, we focus on a sample of nine B[e] stars in different evolutionary phases. 
Emission of the [O\,{\sc i}] lines is one of the characteristics of B[e] stars.
We find that four of the objects display [Ca\,{\sc ii}] line emission: 
for the B[e] supergiants V1478~Cyg and 3~Pup the kinematics obtained from the [O\,{\sc i}] and [Ca\,{\sc ii}] 
line profiles agrees with a Keplerian rotating disc scenario; the forbidden lines of the compact planetary nebula OY~Gem
display no kinematical broadening beyond spectral resolution; the LBV candidate V1429~Aql shows no [O\,{\sc i}] lines, 
but the profile of its [Ca\,{\sc ii}] lines suggests that the emission originates 
in its hot, ionized circumbinary disc.
As none of the B[e] stars of lower mass displays [Ca\,{\sc ii}] line emission, we conclude that these lines
are more likely observable in massive stars with dense discs, supporting and
strengthening the suggestion that their appearance requires high-density 
environments. 
\end{abstract}




\section{Introduction}

Massive stars ($> 8$\,M$_{\odot}$)  play a major role in the evolution of
their host galaxies. Via their stellar winds, they strongly enrich the
interstellar medium with chemically processed material and deposit large
amounts of momentum and energy into their surroundings during their entire
lifetime, before they end their lives in spectacular supernova explosions.
 During their evolution, massive stars can pass through short-lived
phases of enhanced mass loss and mass ejections. Well-known groups in such
transition phases are the luminous blue variables (LBVs), red supergiants 
(RSGs), yellow hypergiants (YHGs), and B[e] supergiants (B[e]SGs). The ejected 
material typically accumulates in either shells, rings, or disc-like structures 
and often veils the central star. Moreover, the conditions within the ejecta,
with respect to temperature and density, favour a rich chemistry, including the
formation of molecules and condensation of dust particles. Consequently, 
these stars display rich emission line spectra, facilitating studies of 
structure and kinematics of the cirumstellar material.

\begin{table*}
\begin{minipage}{0.95\textwidth}
\caption{Target list. Radial velocities $V_{\rm r}$ were derived using all available 
forbidden emission lines.} 
\label{tab:list}     
\begin{tabular}{l @{~~~}l @{~}r c c @{~~}r l l @{~~}c r}
\hline              
Identifier I   & Identifier II  &  HD    & $\alpha$ & $\delta$  & $V$~~~  & Class$^a$        & Sp. type     &   Ref. & $V_{\rm r}$~~~ \\
 &  &   & (J2000.0) & (J2000.0) &  (mag) &   &  &  & (km\,s$^{-1}$) \\
\hline 
V1478~Cyg      & MWC 349A       &        & 20 32 45.53 & +40 39 36.6 & 13.0     & B[e]SG       & B0-B1.5\,I     & (1) & $ -$9 $\pm$ 2 \\ 
3~Pup          & MWC 570        &  62623 & 07 43 48.47 & $-$28 57 17.4 &  3.9     & B[e]SG       & A2.7\,Ib       & (2) &  23 $\pm$ 2 \\ 
CI~Cam         & MWC 84         &        & 04 19 42.13 & +55 59 57.7 & 11.8     & B[e]SG cand      & B0-2\,I        & (3) & $-$51 $\pm$ 2\\ 
V1972~Cyg      & MWC 342        &        & 20 23 03.61 & +39 29 50.1 & 10.6      & B[e]SG cand  & B0-2           & (4) & $-$23 $\pm$ 1\\  
V1429~Aql      & MWC 314        &        & 19 21 33.97 & +14 52 56.9 &  9.9     & B[e]SG cand  & B3\,Ibe        & (5) &  29 $\pm$ 1\\ 
OY~Gem         & MWC 162        &  51585 & 06 58 30.41 & +16 19 26.1 & 11.2     & B[e] cPNe   & OBpe            & (6) &  54 $\pm$ 1\\ 
V743~Mon       & MWC 158        &  50138 & 06 51 33.40 & $-$06 57 59.4 &  6.6     & B[e] uncl   & B6-7\,III-V    & (7) &  38 $\pm$ 1\\ 
BD+23~3183     & IRAS 17449+2320&        & 17 47 03.28 & +23 19 45.4 & 10.0     & B[e] uncl$^b$   & A0             & (8) & $-$17 $\pm$ 3\\ 
HD~281192      & MWC 728        & 281192 & 03 45 14.72 & +29 45 03.2 &  9.8     & B[e] uncl$^b$   & B5 + G8       & (9) &  28 $\pm$ 2\\ 
\hline          
\end{tabular}
\medskip

$^a$ Classification according to the evolutionary state:
B[e]SG (cand) = B[e] supergiant (candidate),  
B[e] cPNe = compact planetary nebula B[e] star,
B[e] uncl = unclassified B[e] star.\\
$^b$ Objects were suggested by \citet{2007ApJ...671..828M} to be of FS CMa type.

References:
(1)~\citet{2012A&A...541A...7G};
(2)~\citet{2015AstBu..70...99K};
(3)~\citet{2002A&A...392..991H}; 
(4)~\citet{1999A&A...349..126M}; 
(5)~\citet{2010A&A...517A..67C};
(6)~\citet{1996A&AS..117..281J};
(7)~\citet{2009A&A...508..309B}; 
(8)~\citet{2007ApJ...671..828M}; 
(9)~\citet{2015ApJ...809..129M}.
\end{minipage}
\end{table*}

\begin{table*}
\begin{minipage}{0.94\textwidth}
\caption{Observing log.}  
\label{tab:log}      
\begin{tabular}{l c r r c r r c r r}
\hline              
Star           &\multicolumn{3}{c}{H~$\alpha$}&\multicolumn{3}{c}{Ca IR triplet}   & \multicolumn{3}{c}{[Ca\,{\sc ii}]}     \\
               & Date   &  $T_{\rm exp}$ (s)  & S/N  &   Date       &  $T_{\rm exp}$ (s)  &   S/N &   Date         &  $T_{\rm exp}$ (s)   &   S/N \\
\hline             

V1478~Cyg      & 2012-08-15 &  3600~~~ &  25  &   2012-08-15 &   3600~~~   &   35  &   2012-08-15  &  3600~~~  &   35   \\
3~Pup          & 2011-10-16 &   400~~~ &  135 &   2012-04-07 &    370~~~   &  45   &   2011-10-16  &   400~~~  &  150  \\        
CI~Cam         & 2011-10-01 &  3600~~~ &   60 &   2011-09-30 &   3600~~~   &  40   &   2011-09-30  &  3600~~~  &   50   \\
V1972~Cyg      & 2011-08-19 &  3001~~~ &   80 &   2012-04-30 &   3600~~~   &  60   &   2011-08-18  &  3600~~~  &  100   \\
V1429~Aql      & 2012-03-24 &  3600~~~ &   65 &   2012-03-24 &    707~~~   &  25   &   2012-03-24  &  3600~~~  &   50  \\
OY~Gem         & 2012-03-20 &  3600~~~ &   17 &   2012-03-18 &   3169~~~   &   8   &   2012-03-18  &  1347~~~  &  10   \\
V743~Mon       & 2012-01-31 &  2700~~~ &  135 &   2012-01-31 &   3600~~~   &  80   &   2012-02-03  &  3600~~~  &  140   \\
BD+23~3183     & 2011-08-23 &  3600~~~ &   80 &   2012-08-15 &   3600~~~   &  50   &   2011-08-23  &  3600~~~  &   80   \\
HD~281192      & 2012-08-20 &  3600~~~ &  100 &   2012-08-16 &   3600~~~   &  55   &   2012-08-16  &  3600~~~  &   80   \\
\hline      
\end{tabular}
\end{minipage}
\end{table*}

Besides the lines from the hydrogen recombination series, forbidden emission
lines belong to the prominent features observed from circumstellar
environments. The shape and strength of their profiles provide direct
access to the structure (density, temperature) and gas dynamics of their
formation regions. The lines of [O\,{\sc i}] $\lambda\lambda$6300, 6364
and [Ca\,{\sc ii}] $\lambda\lambda$7291, 7324 have recently proven to trace
predominantely high-density gas regions, such as the innermost disc regions
of B[e]SGs \citep{2007A&A...463..627K, 2010A&A...517A..30K,
2012MNRAS.423..284A}. They were also reported from a few young stellar objects
\citep{1994ApJS...93..485H, 2004ApJ...609..261H}, as well as from the dense
winds or outflows of YHGs \citep[e.g.,][]{2013ApJ...773...46H,
2014ApJ...790...48H}. Other forbidden lines, such as [Fe\,{\sc ii}],
[S\,{\sc ii}], [N\,{\sc ii}], etc., trace regions of low(er) density but
diverse temperatures and deliver complementary information from physically
disjoint regions. The permitted lines of the Ca\,{\sc ii} IR triplet
($\lambda\lambda$8498, 8542, 8662) are observed in a variety of objects,
including T\,Tauri stars \citep[e.g.,][]{1992ApJS...82..247H,
2011MNRAS.411.2383K}, Herbig Ae/Be stars \citep[e.g.,][]{1992ApJS...82..285H},
classical Be stars \citep*[e.g.,][]{1976IAUS...70...59P, 1981A&A...103....1B,
1988A&A...192..285J, 1988A&AS...72..129A, 1990A&AS...84...11A}, B[e]SGs
\citep{2010A&A...517A..30K, 2012MNRAS.423..284A}, and other, non-supergiant
B[e] stars \citep{2001ApJS..136..747B, 2009A&A...508..309B}. These triplet
lines are often composite, possessing contributions from both a less dense wind
and a dense shell or disc \citep{2012MNRAS.423..284A}.

Interestingly, the lower energy levels, to which the calcium  infrared (IR) triplet
lines decay, are the upper levels of two forbidden transitions ([Ca\,{\sc ii}]
$\lambda\lambda$7291, 7324). Therefore, the appearance of the [Ca\,{\sc ii}]
lines could be a suitable tracer for specific temperature and density
conditions within the high-density environments of emission-line stars.
Pioneering investigations have recently been started by
\citet{2012MNRAS.423..284A} for the discs of B[e]SGs. In their studies, the
authors found that the [Ca\,{\sc ii}] lines in these objects originate from
disc regions close to the star, and the
[O\,{\sc i}] lines  arise from either the same or a slightly lower density region,
i.e., from farther out. As the discs of B[e]SGs are typically in Keplerian
rotation \citep[see, e.g.,][]{2011A&A...526A.107M, 2012A&A...548A..72C,
2012A&A...543A..77W, 2010A&A...517A..30K, 2013A&A...549A..28K,
2014ApJ...780L..10K, 2015ApJ...800L..20K, 2015AJ....149...13M}, the
analysis of the profiles of both the [Ca\,{\sc ii}] and [O\,{\sc i}] lines
have been used by \citet{2012MNRAS.423..284A} to constrain the kinematics and
the inclination angles. These results, combined with the outcome of earlier
studies \citep[see][]{2005A&A...441..289K}, clearly demonstrate the great
potential of the different, complementary sets of forbidden emission lines in
determining structure and kinematics of circumstellar gas.

However, the frequency and origin of the [Ca\,{\sc ii}] lines in stars with
circumstellar discs or dense environments is not well studied yet. Also unclear
are the proper physical conditions in terms of temperature and density ranges
 necessary for the excitation of these forbidden lines. Hence, systematic surveys
and studies of stars with 
high-density environments are needed to shed  light 
on a possible link between the 
appearance of the [Ca\,{\sc ii}] lines and other prominent disc 
emission features, such as the [O\,{\sc i}] and the Ca\,{\sc ii} IR triplet 
lines, and to provide important information on the physical conditions in their formation regions. 
Therefore, we started a survey of emission-line stars and disc sources 
in a variety of evolutionary stages, aimed at finding objects that show the 
strategic [Ca\,{\sc ii}] and [O\,{\sc i}] lines as possible tracers of 
high-density discs or shells.

One group of stars surrounded by dusty shells, rings or discs, as is evident
from their infrared excess emission \citep{1976A&A....47..293A}, are the
B[e] stars (or, more precisely, stars showing the B[e] phenomenon). The
optical spectra of these objects typically display strong Balmer line emission,
together with emission of low-ionized metals from both permitted and forbidden
transitions. These spectral characteristics are seen in stars of diverse
evolutionary phases. \citet{1998A&A...340..117L} were the first to sort B[e]
stars according to their evolutionary stage. They found pre-main sequence
(Herbig Ae/B[e]) stars, as well as evolved 
objects such as compact planetary nebulae and B[e]SGs.
In addition, several B[e] stars
turned out to be symbiotic objects. However, for more than half of the known
B[e] stars no appropriate evolutionary stage could be assigned, either because
of the lack of proper stellar parameters, such as distance hence luminosity, or
because the stars display spectral characteristics in common with more than
just one classification. These stars are gathered within the group of
unclassified B[e] stars.


\begin{table*}
\vskip -10pt
\begin{minipage}{0.94\textwidth}
\caption{Shapes (Sh), peak separations (PS, in km\,s$^{-1}$), full widths at zero
 intensity (ZI, in km\,s$^{-1}$) and full widths at half maximum 
 (HM, in km\,s$^{-1}$) of forbidden oxygen and calcium lines. Errors are shown as superscript.
 Classification of the line profile types: 
 S -- single peaked emission line;
 D -- double peaked emission line;
 W -- weak emission line with unclear (not resolved) profile shape;
 X -- no line detected.
}
\label{tab:forbidden}
\begin{tabular}{@{\extracolsep{-1pt}}@{}l*{17}{c}@{}}
\hline 
Object & \multicolumn{4}{c}{[O\,{\sc i}] 6300 \AA}&
\multicolumn{4}{c}{[O\,{\sc i}] 6364 \AA}
& \multicolumn{4}{c}{[Ca\,{\sc ii}] 7291 \AA}&
\multicolumn{4}{c}{[Ca\,{\sc ii}] 7324 \AA}\\
               & Sh & PS      & ZI         & HM         & Sh& PS & ZI        & HM       & Sh& PS       & ZI         & HM       & Sh& PS       & ZI  & HM \\
\hline  
V1478~Cyg      & S  & --      & 150$^{10}$ &  63$^{2}$  & S & -- &160$^{15}$ &68$^{2}$  & D &n.r.      & 150$^{20}$ & 80$^{5}$ & D & 41$^{1}$ & 140$^{20}$&90$^{5}$   \\
3~Pup          & D  &48$^{5}$ & 165$^{10}$ & 105$^{5}$  & W & -- &190$^{25}$ &93$^{10}$ & D & 65$^{8}$ & 147$^{15}$ & 105$^{8}$& D & 56$^{5}$ & 150$^{15}$&  95$^{8}$ \\
CI~Cam         & X  & --      &  --        & --         & X & -- & --        & --       & X & --       & --         & --       & X & --       & --        & --        \\
V1972~Cyg      & S  & --      & 143$^{5}$  &45.6$^{1}$  & S & -- & 144$^{5}$ &45.5$^{1}$& X & --       & --         & --       & X & --       & --        & --        \\     
V1429~Aql$^a$  & W  & --      & 140$^{20}$ &  61$^{5}$  & X & -- & --        & --       & D & 61$^{2}$ & 156$^{5}$  & 105$^{3}$& D & 62$^{2}$ & 146$^{5}$ & 109$^{2}$ \\
OY~Gem         & S  & --      & 120$^{10}$ &  39$^{2}$  & S & -- &115$^{10}$ &37$^{2}$  & S & --       & 80$^{10}$  & 27$^{1}$ & S & --       & 70$^{5}$  & 24$^{1}$  \\
V743~Mon       & S  & --      & 156$^6$    &  52$^1$    & S & -- & 142$^{13}$& 51$^1$   & X & --       & --         & --       & X & --       & --        & --        \\ 
BD+23~3183     & S  & --      & 150$^{10}$ & 120$^{10}$ & S & -- & 120$^{10}$& 41$^5$   & X & --       & --         & --       & X & --       & --        & --        \\
HD~281192$^{b}$& S  & --      & 160$^{10}$ &  78$^{1}$  & S & -- & 170$^{20}$&80$^{2}$  & X & --       & --         & --       & X & --       & --        & --        \\
\hline      
\end{tabular}
$^a$Identification of the [O\,{\sc i}] 6300 {\AA} line uncertain.
$^b$Indication of a slightly (but unresolved) double-peaked profile of the [O\,{\sc i}] lines. 
\end{minipage}
\end{table*}

\section{Observations and data reduction}
\label{sec:obs}

Our sample contains nine Galactic northern B[e] stars 
of different masses and in different evolutionary stages. In particular, we have
three supergiants, two supergiant candidates, one compact planetary nebula,
and three B[e] stars with unclear evolutionary state. The objects
are listed in Table\,\ref{tab:list} together with their coordinates, $V$ band
magnitudes, classification, and spectral types collected from the literature.

The observations were obtained using the Coud\'{e} spectrograph attached
to the Perek 2-m telescope at Ond\v{r}ejov Observatory \citep{2002PAICz..90....1S}.
We used the 830.77 lines mm$^{-1}$ grating produced by Bausch \& Lomb with a SITe 2030$\times$800 CCD.
Spectra were taken in three different wavelength regions:
around H~$\alpha$ (6250\,\AA \ to  6760\,\AA), in the region of the [Ca\,{\sc ii}] $\lambda\lambda$7291,
7324 lines (6990\,\AA \ to  7500\,\AA), and in the region of the Ca\,{\sc ii} IR triplet (8470\,\AA \ to  8980\,\AA). 
The H~$\alpha$ region also encloses the two [O\,{\sc i}] $\lambda\lambda$6300, 6364
lines. The spectral resolution achieved in each range is $R\simeq$ 13\,000 in
the H~$\alpha$ region, $R\simeq$ 15\,000 in the
[Ca\,{\sc ii}] region, and $R\simeq$ 18\,000 in the
Ca\,{\sc ii} IR triplet region. For
wavelength calibration, a comparison spectrum of a ThAr lamp was taken
immediately after each exposure. The stability of the wavelength scale was
verified by measuring the wavelength centroids of [O\,{\sc i}] sky lines. The
velocity scale remained stable within 1\,km\,s$^{-1}$.

Spectra in the three different wavelength regions were taken within the same
night or, if this was not possible, close in time. Details on the observations
and the quality of the spectra are provided in the observing log
(Table\,\ref{tab:log}). The data were reduced using standard
IRAF\footnote{IRAF is distributed by the National Optical Astronomy
Observatories, which are operated by the Association of Universities for
Research in Astronomy, Inc., under cooperative agreement with the National
Science Foundation.} tasks, such as bias subtraction, flat-field normalization,
and wavelength calibration. To perform telluric corrections in all three
wavelength regions, telluric standard stars were observed each night. The final
spectra were corrected for heliocentric and systemic velocities and normalized.
The systemic velocities were obtained from the forbidden lines in our spectra,
and the values are included in Table\,\ref{tab:list}.

\section{Results}
\label{sec:results}

\begin{figure*}
\includegraphics[scale=1.2]{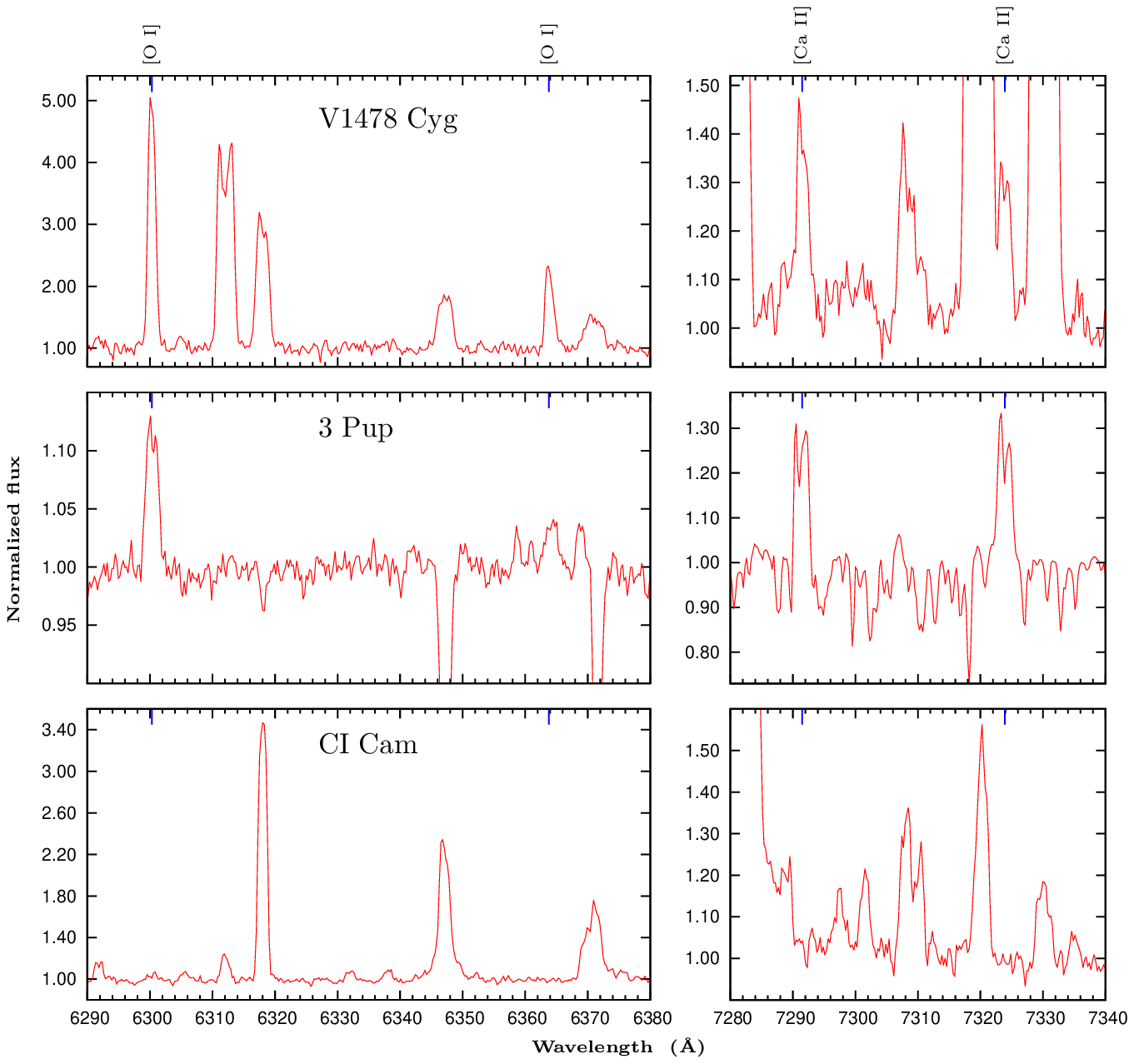}
\caption{Spectral portions of B[e]SG stars covering the strategic forbidden lines of [O\,{\sc i}] $\lambda\lambda$6300, 6364 and [Ca\,{\sc ii}]
$\lambda\lambda$7291, 7324. The wavelengths of the lines are marked by ticks. Spectra are corrected for heliocentric and systemic velocities.}
\label{fig1v2}
\end{figure*}

\begin{figure*}
\includegraphics[scale=1.2]{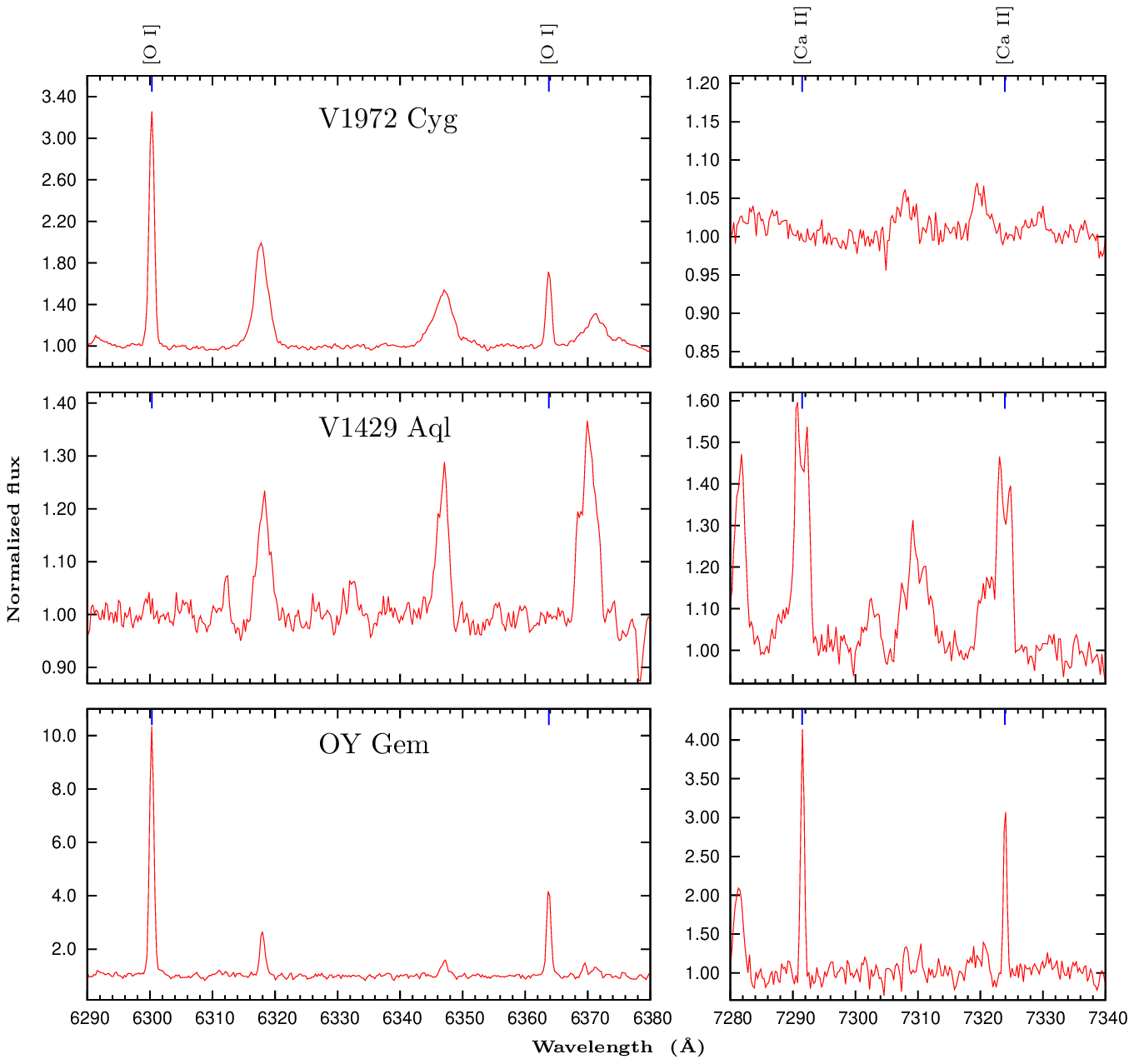}
\caption{As Fig.\,\ref{fig1v2}, but for the two B[e]SG candidates and the 
compact planetary nebula.}
\label{fig2v2}
\end{figure*}

\begin{figure*}
\includegraphics[scale=1.2]{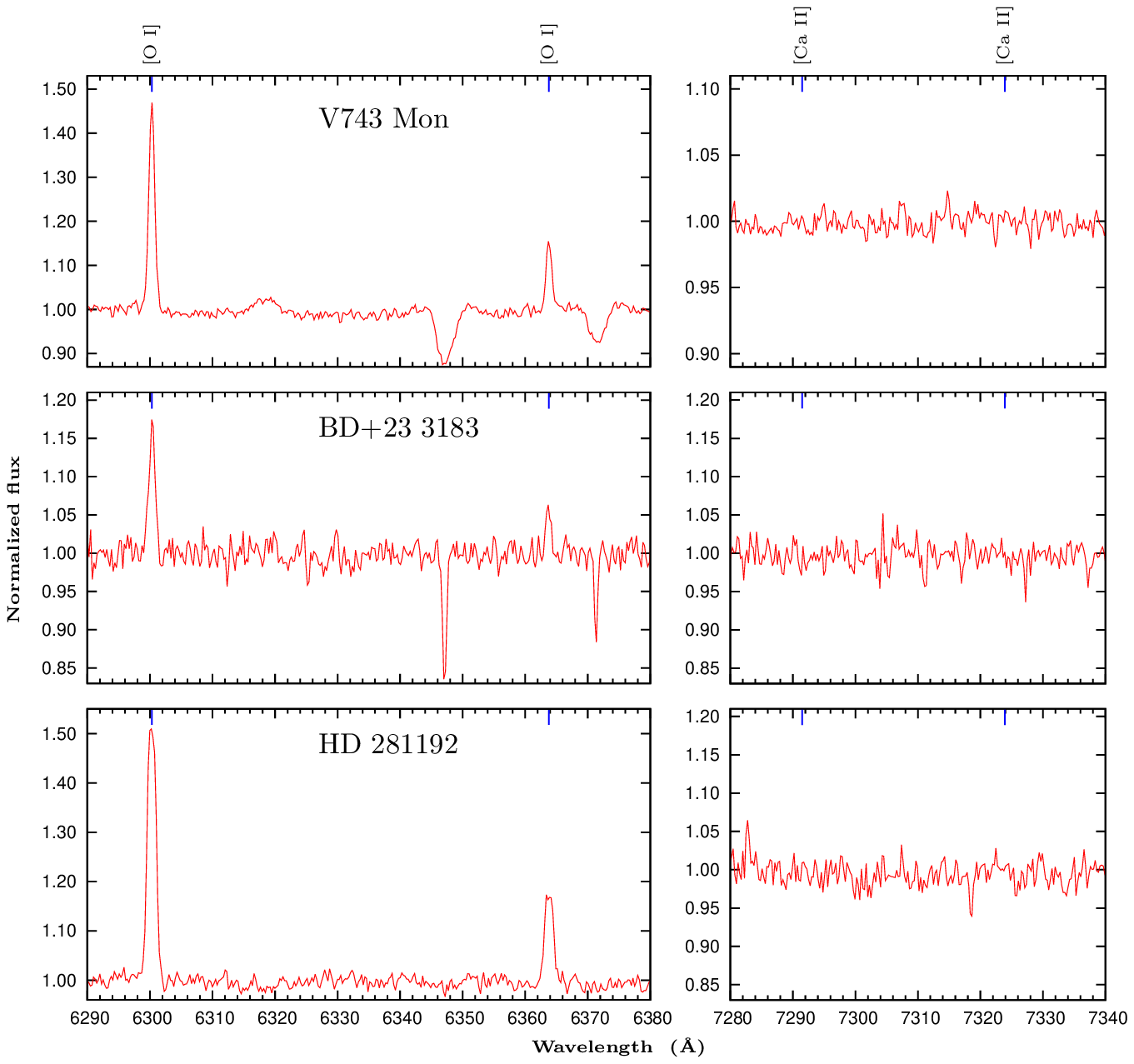}
\caption{As Fig.\,\ref{fig1v2}, but for the unclassified B[e] stars.}
\label{fig3v2}
\end{figure*}

\begin{figure*}
\includegraphics[scale=1.2]{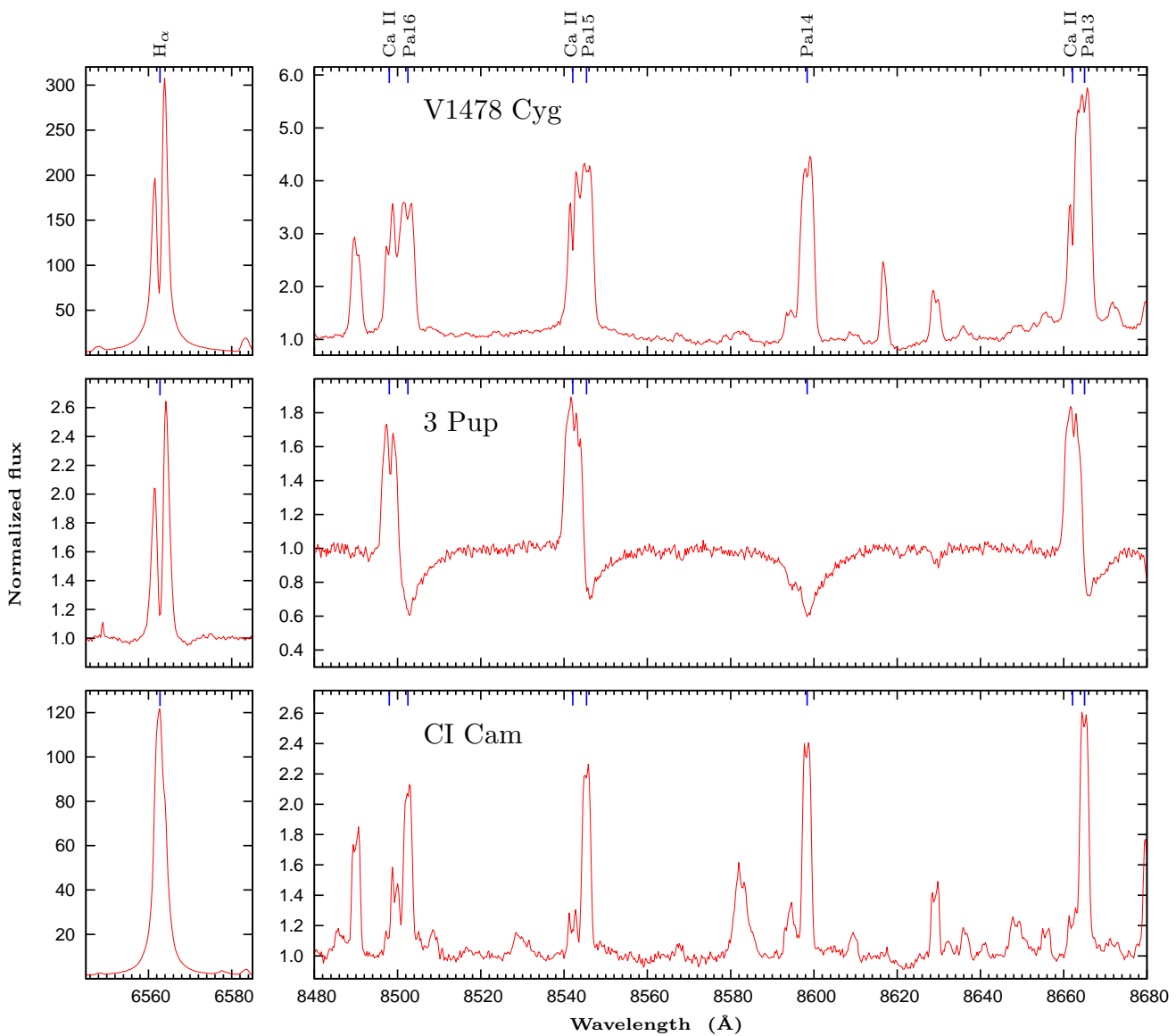}
\caption{Spectral portions of B[e]SG stars covering the strategic lines of 
H~$\alpha$ and the Ca\,{\sc ii} $\lambda\lambda$8498, 8542, 8662 IR triplet. 
The wavelengths of the lines (including the four Paschen lines) are marked by 
ticks. Spectra are corrected for heliocentric and systemic velocities.}
\label{fig4v2}
\end{figure*}

\begin{figure*}
\includegraphics[scale=1.2]{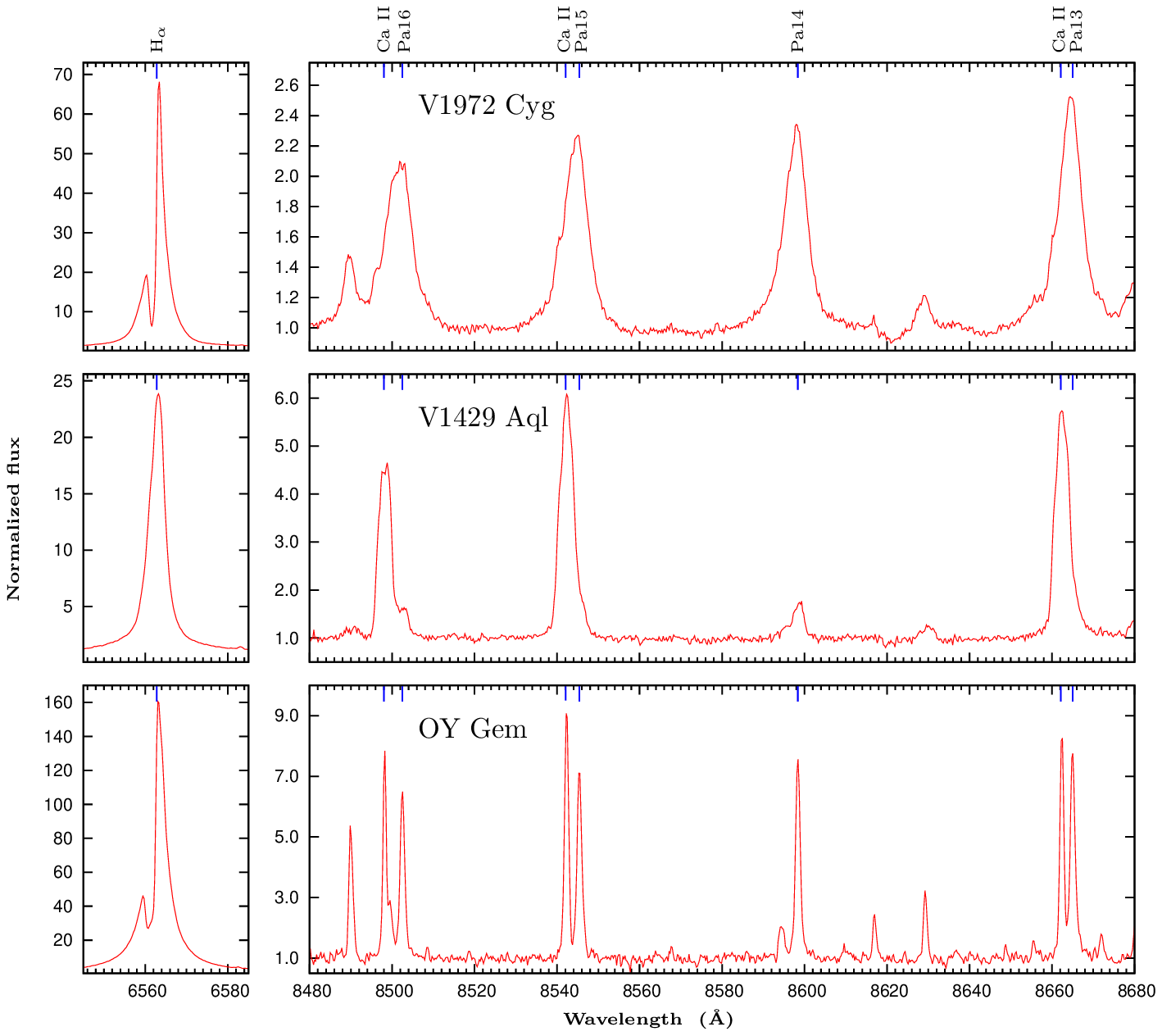}
\caption{As Fig.\,\ref{fig4v2}, but for the two B[e]SG candidates and the 
compact planetary nebula.}
\label{fig5v2}
\end{figure*}

\begin{figure*}
\includegraphics[scale=1.2]{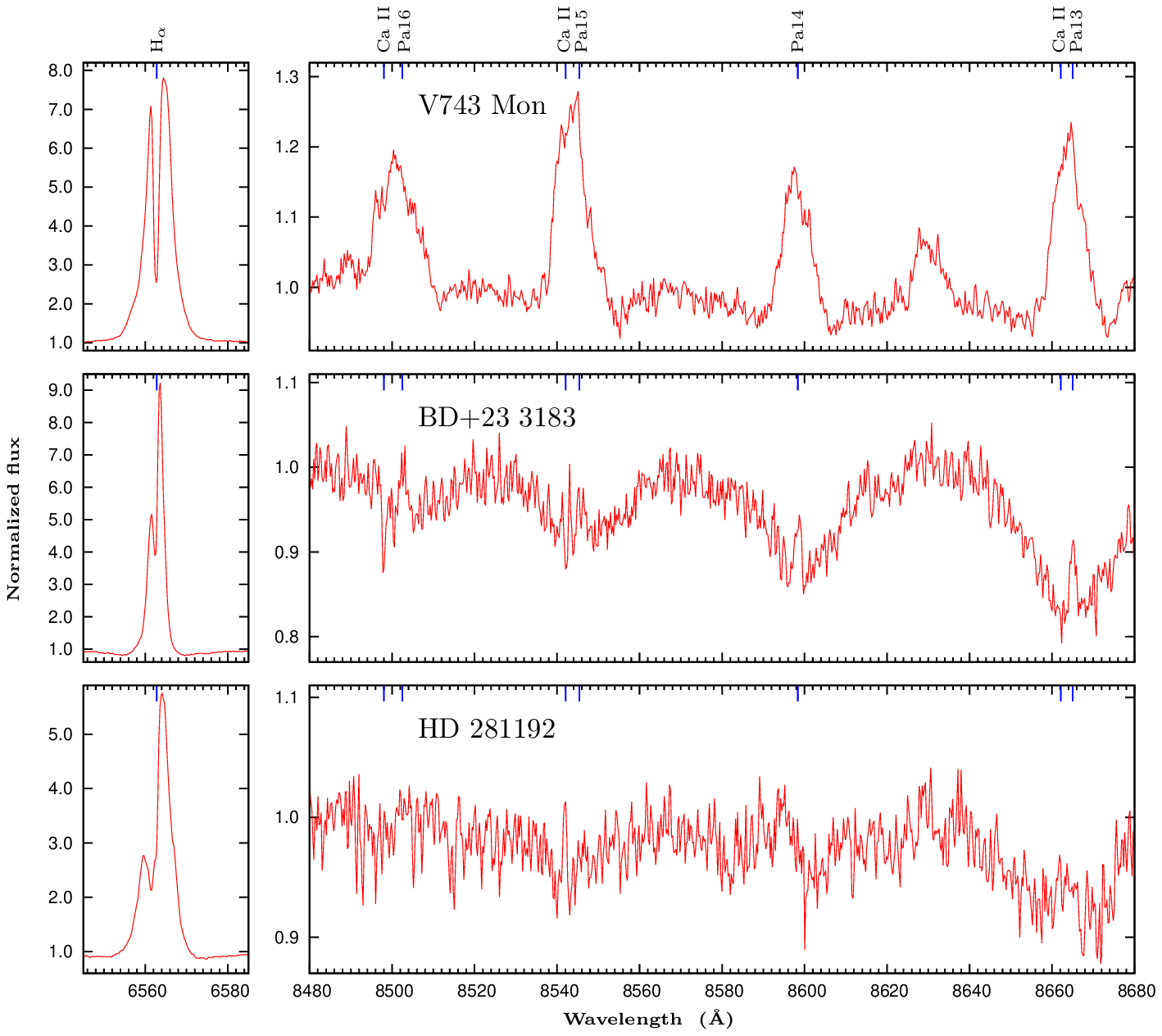}
\caption{As Fig.\,\ref{fig4v2}, but for the unclassified B[e] stars.}
\label{fig6v2}
\end{figure*}

Figs.\,\ref{fig1v2}-\ref{fig3v2} display the spectral ranges around the 
forbidden lines of our interest. Shown are the regions of the [O\,{\sc i}] 
$\lambda\lambda$6300, 6364 lines (left) and of the [Ca\,{\sc ii}] 
$\lambda\lambda$7291, 7324 lines (right). We grouped the B[e]SGs in 
Fig.\,\ref{fig1v2}, the B[e]SG candidates and the compact planetary nebula in 
Fig.\,\ref{fig2v2}, and the remaining objects in Fig.\,\ref{fig3v2}.
Maintaining the same order of the stars, we show in 
Figs.\,\ref{fig4v2}-\ref{fig6v2} the H~$\alpha$ 
line (left) and the Ca\,{\sc ii} IR triplet ($\lambda\lambda$8498, 8542, 
8662, right). The latter region also encompasses four hydrogen Paschen (Pa) 
lines, Pa(16) to Pa(13). Three of them appear in the close vicinity
of the Ca\,{\sc ii} triplet lines. In most of our spectra, Ca\,{\sc ii} triplet and 
Paschen lines are blended, hampering a clear identification and classification
of their profile shapes. The wavelengths of all these lines are marked in
each panel and their identification is given on top of the figures.

\begin{table}
\caption{Shapes of H~$\alpha$ and
Ca\,{\sc ii} IR triplet lines. Classification of the line profile types:
 S -- single peaked emission line;
 D -- double peaked emission line;
 B -- emission line strongly blended with Pa line, no classification possible;
 DWA -- double peaked emission line on top of wide absorption component;
 X -- no line detected.
 }              
\label{tab:IR}      
\begin{tabular}{l c c c c c}
\hline
Object & H~$\alpha$& Ca\,{\sc ii} & Ca\,{\sc ii}& Ca\,{\sc ii}\\
       &           & 8498 \AA     & 8542 \AA    & 8662 \AA\\
\hline
V1478~Cyg      & D    & D  & D   & D   \\
3~Pup          & DWA  & D  & D   & D   \\
CI~Cam         & S    & D  & D   & D   \\
V1972~Cyg      & D    & B$^a$  & B$^a$   & B$^a$   \\
V1429~Aql      & S    & S  & S   & S   \\
OY~Gem         & D    & S  & S   & S   \\
V743~Mon       & D    & B  & B   & B   \\
BD+23~3183     & DWA  & X  & X   & X   \\
HD~281192      & DWA  & X  & X   & X   \\
\hline
\end{tabular} 

$^a$Identification uncertain.
\end{table}

Inspection of the spectra reveals that all objects display [O\,{\sc i}] emission, 
except CI~Cam and, possibly, V1429~Aql. Although V1429~Aql has a weak 
emission feature in the vicinity of the wavelength of the [O\,{\sc i}] 
$\lambda$6300 line, its radial velocity of only $\sim$\,5\,$\pm$\,10\,km\,s$^{-1}$ 
disagrees with the value measured from other emission lines in that star 
(29\,$\pm$\,1\,km\,s$^{-1}$, see Table\,\ref{tab:list}). The two 
objects have further in common that their H~$\alpha$ line is single peaked, 
while in all other stars H~$\alpha$ is double-peaked, with the red peak 
typically more intense than the blue one. 


\begin{table*}
\begin{minipage}{0.94\textwidth}
\caption{Stellar parameters of the sample stars.}
\label{tab:stel}
\begin{tabular}{l c c c c c c l}
\hline
Star           & $T_{\rm eff}$       & $\log L/$L$_{\odot}$ & $M_{*}$       & Disc    & $i^{a}$    & CO Bands & Reference  \\
               & [K]                 &                    & [M$_{\odot}$] &         & [\degr]    &          &  \\
\hline
V1478~Cyg      & 24\,000 $\pm$ 4\,000 & 5.75 $\pm$ 0.15     & 38            & yes     & 82         & emis      & (1), (2), (3), (4) \\
3~Pup          & 8\,250 $\pm$ 250    & 4.33 $\pm$ 0.09     & 15-20         & yes     & 38         & emis      & (5), (6)  \\
CI~Cam         & 24\,000 $\pm$ 6\,000 & $>$ 5.4            & $>$ 12        & possibly$^b$ & $\sim$ 0& emis      & (7), (8)  \\
               & 20\,000 $\pm$ 2\,000 & $<$ 4.0            &               & uncertain &            & no       & (9), (10)  \\
V1972~Cyg      &  26\,000            & 4.2 $\pm$ 0.4       & --            & uncertain     & --         & no       & (11), (12), (13)  \\
V1429~Aql      & 18\,000           & 5.85               & 66 $\pm$ 9$^c$      & yes     & 73 $\pm$ 13 & no       & (14), (15), (10)  \\
OY~Gem         & 28\,000             & 3.8                & 0.62          & uncertain      & --         & no       & (16), (17)  \\
V743~Mon       & 13\,200 $\pm$ 500    & 3.06 $\pm$ 0.27     & 5             & yes     & 56 $\pm$ 4  & no       & (18), (19), (6)  \\
BD+23~3183     & 10\,000             &                    & $<$ 4         & uncertain & --         & unknown  & (20)  \\
HD~281192      & 14\,000         &                    & 4 $\pm$ 0.5       & uncertain & --         & abs      & (21), (22)  \\
\hline
\end{tabular}

$^{a}$ Inclination angle of the system rotation axis with respect to the line of sight.
Inclination of 90{\degr} means edge-on view of the disc. \\
$^b$ During the 
outburst phase.
$^c$ Total mass of the binary system.

References: 
(1)~\citet{2002A&A...395..891H}
(2)~\citet{2013A&A...553A..45B};
(3)~\citet{1987ApJ...312..297G};
(4)~\citet{2000A&A...362..158K};
(5)~\citet{2011A&A...526A.107M}; 
(6)~\citet{2012BAAA...55..123M};
(7)~\citet{2002A&A...392..991H}; 
(8)~\citet{1999A&A...348..888C};
(9)~\citet{2002A&A...390..627M};
(10)~\citet{2014MNRAS.443..947L};
(11)~\citet{1999A&A...349..126M};
(12)~\citet{2005MNRAS.364..725O};
(13)~M.\,L.~Arias, private communication;
(14)~\citet{2013A&A...559A..16L};
(15)~\citet{2008A&A...487..637M}; 
(16)~\citet{1992SvAL...18..418A}; 
(17)~\citet{2008MNRAS.385.1076V};
(18)~\citet{2009A&A...508..309B}; 
(19)~\citet{2011A&A...528A..20B}; 
(20)~\citet{2007ApJ...671..828M}; 
(21)~\citet{2015ApJ...809..129M}; 
(22)~Arias~et~al. in preparation.

\end{minipage}

\end{table*}

The [Ca\,{\sc ii}] lines are
detected in four objects: V1478~Cyg, 3~Pup, V1429~Aql, and OY~Gem. All stars
with [Ca\,{\sc ii}] emission also have emission in the Ca\,{\sc ii} IR
triplet lines. The reverse is, however, not true, with CI~Cam, V1972~Cyg,
and V743~Mon as examples. In the latter two stars the triplet lines are
strongly blended with broad Paschen line emission that dominates the red
spectra. In contrast, the triplet lines are prominent in 3~Pup and V1429~Aql,
in which the Pa lines are either in absorption or present only weak emission
features, respectively. Two of the objects, BD+23~3183 and HD~281192, display
emission of neither the [Ca\,{\sc ii}] nor the IR triplet lines. Both
stars are unclassified regarding their evolutionary phase, but are presumably
dwarfs with masses $< 8$\,M$_{\odot}$ \citep{2007ApJ...671..828M,2015ApJ...809..129M}.

Where appropriate, we measure the width of the lines in velocity units at
the continuum level and at half of the maximum value, and for the
double-peaked lines also the peak separation. These values,
together with a description of the shape of the individual lines, are listed in
Table\,\ref{tab:forbidden} for the forbidden lines of [O\,{\sc i}] and
[Ca\,{\sc ii}]. For H~$\alpha$ and the Ca\,{\sc ii} IR triplet, we only
provide details on their profile shape (if resolved) in Table\,\ref{tab:IR}.
Moreover, stellar parameters were collected from the literature and
summarized in Table\,\ref{tab:stel}. In this table we also provide information
on the existence of a circumstellar disc and on emission of CO bands, which
could mark the inner rim of the molecular disc. We mark the presence of a disc only for those objects with yes, for which a definite proof for the existence is available, either directly from (spectro-)polarimetric or inter\-fero\-met\-ric observations, or indirectly from kinematical analyses of a large number of emission lines. 
Although a double-peaked H~$\alpha$ line profile is a strong indication for a gas disc, this alone cannot be considered as a proof of a disc-like structure, because such a profile can arise in a variety of configurations of the circumstellar matter, even in a spherically symmetric wind, as shown by \citet{1993ApJ...411..874C}.
Hence, stars that were reported so far only as having a double-peaked H~$\alpha$ line profile are marked in the table with uncertain disc presence. 
In the following, we review
each object and discuss the observed spectral appearance in the frame of the
characteristics of the circumstellar environment.

\subsection*{V1478~Cyg = MWC~349A}

V1478~Cyg (MWC~349A) is the more massive component of MWC~349. The B component is a 
B0\,III star, located 2.4$\arcsec$ west of the main component, which is
a highly reddened ($A_{\rm V} \simeq 10$\,mag) B0-1.5\,I star. Luminosity 
determinations for the main component range from $\sim 3\times 
10^{4}$\,L$_{\odot}$ \citep{1985ApJ...292..249C} to $(4-8)\times 
10^{5}$\,L$_{\odot}$ \citep{2012A&A...541A...7G}. V1478~Cyg is one of the 
brightest radio sources, and its emission at 2\,cm shows an hourglass-shaped 
bipolar nebula \citep{1985ApJ...297..677W}.

Besides a high-density ionized wind, V1478~Cyg displays strong infrared excess 
emission, and near-infrared speckle interferometry resolved a disc-like dust 
structure oriented in east-west direction, seen nearly edge-on 
\citep{1986A&A...155L...6L, 1983A&A...120..237M}. V1478~Cyg is unique in the 
sense that it is the only object with hydrogen recombination maser line 
emission in the mm and sub-mm range \citep[e.g.,][]{1989A&A...215L..13M}. 
These masers originate from a dense gas disc, and the kinematics, obtained
from their line profiles, imply Keplerian rotation of the disc
(\citealp*{1992A&A...256..507T}, \citealp{1994A&A...283..582T}), as well as 
a rotating wind emanating from the disc \citep{2011A&A...530L..15M,
2013A&A...553A..45B}. Furthermore, CO band emission has been detected 
\citep{1987ApJ...312..297G, 2000A&A...362..158K}. The latter authors suggested 
that this emission originates from the far side of the inner molecular edge of 
the Keplerian disc. 

\citet{2008A&A...477..193M} discovered a thin H~$\alpha$ shell of approximately 
2.5$\arcmin$ diameter north of V1478~Cyg and a parsec-scale hourglass-shaped 
nebula is also seen in the mid-infrared \citep{2012A&A...541A...7G, 
2013ApJ...777...89S}. The observed characteristics favour an evolved nature of
the star \citep{2002A&A...395..891H, 2012A&A...541A...7G}. However, a pre-main 
sequence evolutionary stage cannot be ruled out, based on the recently 
discovered proximity of V1478~Cyg to a molecular cloud with identical radial 
velocity \citep{2013ApJ...777...89S}.

V1478~Cyg displays a pure emission line spectrum in the wavelength ranges 
covered by our observations. With a line intensity of about 300 times the 
continuum, V1478~Cyg is the strongest H~$\alpha$ emitting source of our sample. 
The line profile is double-peaked with the red peak about twice the intensity of the 
blue peak. The Paschen lines are narrow and slightly double-peaked. They are 
blended with the Ca\,{\sc ii} IR triplet lines, which are also 
double-peaked alike the [Ca\,{\sc ii}] lines. Only the [O\,{\sc i}] lines show 
single-peaked profiles, but have been reported by \citet{2003A&A...408..257Z} 
to be slightly split as well when observed with (much) higher resolution. The 
diversity in the line profiles of the forbidden lines would be in agreement 
with the B[e]SG disc scenario of \citet{2012MNRAS.423..284A} in which the 
[O\,{\sc i}] lines formed farther away from the star (at smaller rotational 
velocity) than the [Ca\,{\sc ii}] lines.


\subsection*{3~Pup = MWC~570 = HD~62623}

3~Pup is so far the only known Galactic A[e] supergiant. 
Emission from the Ca\,{\sc ii} IR triplet and the [Ca\,{\sc ii}] lines 
was seen by \citet{1947ApJ...105..212H} and \citet{1994PASP..106..382D}, 
although the latter authors mistakenly identified the [Ca\,{\sc ii}] lines as 
the [O\,{\sc ii}] doublet lines. Based on radial velocity variations, 
\citet*{1995A&A...293..363P} proposed that 3~Pup could be an interacting 
binary surrounded by gas and a dusty disc. Their analysis revealed an orbital 
period of either 131 or 161 days.

No remarkable deviation from spherical symmetry of the dusty envelope was 
noticed by \citet{2004ApJ...605..436M} implying that the disc is probably seen
close to pole-on. Interferometric observations with VLTI/MIDI 
\citep{2010A&A...512A..73M} and VLTI/AMBER \citep{2011A&A...526A.107M} resolved 
the disc and constrained the inclination angle to 38$\degr$. The disc consists 
of an inner, ionized gaseous part and an outer dusty part. This disc seems to 
be detached from the central star, as no emission was detected from the innermost region
\citep{2011A&A...526A.107M}. This gap between star and disc could be caused by 
a close companion, as suggested from radial velocity 
variability \citep{1995A&A...293..363P}. However, interferometry 
revealed no indication for such a companion \citep{2011A&A...526A.107M}. A
double-peaked emission profile of Br~$\gamma$ was resolved by AMBER and
clearly attributed to rotation of the gas disc \citep{2011A&A...526A.107M}.
Double-peaked profiles were also reported for the optical emission lines from 
both permitted and forbidden transitions \citep*{2010AstBu..65..150C}, as well 
as from CO band and SiO band emission in the near-infrared region 
\citep{2012ASPC..464...67M, 2015ApJ...800L..20K}, supporting the Keplerian 
rotating disc scenario.

Our spectra show that all strategic lines ([O\,{\sc i}], [Ca\,{\sc ii}], 
Ca\,{\sc ii} IR triplet, and H~$\alpha$) are in emission with 
double-peaked profiles. Remarkably, H~$\alpha$ has the least intensity of all 
our sample stars, supporting the results from the infrared spectra, i.e. a lack
of gas in the vicinity of the star. Its narrow emission profile is superimposed 
on a wider absorption profile. The peak separation of $\sim$120\,km\,s$^{-1}$ measured
from our H~$\alpha$ profile agrees with the velocity obtained for Br~$\gamma$ \citep{2011A&A...526A.107M}.
Hence, both lines should originate from the same region, i.e., from the detached gas ring resolved by interferometry \citep{2011A&A...526A.107M}.
The Paschen lines are in pure absorption.
Their broad wings are blended with the emission from the Ca\,{\sc ii} IR 
triplet lines. The appearance of both sets of forbidden emission lines 
indicates a high density within the gaseous disc of 3~Pup.


\subsection*{CI~Cam = MWC~84}

The star CI~Cam was identified as the optical counterpart of the X-ray source 
XTE\,J0421+560. It is an eccentric binary system consisting of a bright optical 
object and a compact companion, possibly a white dwarf 
\citep*[e.g.,][]{2004ApJ...601.1088I}. \citet{2006ARep...50..664B} derived a 
period of 19.41 days and an eccentricity of 0.62. Furthermore, they suggested 
the primary to be a massive ($>$\,12\,M$_{\odot}$) star.

CI~Cam became famous due to its spectacular outburst in 1998 over the complete 
wavelength range, from $\gamma$-ray to radio, during which the object 
brightened by 2-3 magnitudes. At that time, optical 
\citep*[e.g.,][]{2002ApJ...565.1169R} and infrared spectroscopy 
\citep[e.g.,][]{1999A&A...348..888C} revealed a rich emission line spectrum 
that changed drastically with time, indicating a highly dynamical environment, 
dense enough to hide the central star. Most of the emission lines presented 
triple-peaked profiles, suggesting a gas disc seen under an intermediate 
inclination angle, and an additional dense, probably ring-shaped region farther 
away from the star \citep{2002A&A...390..627M}. Strong near- and mid-infrared 
excess emission 
due to circumstellar dust was reported by \citet{1973MNRAS.161..145A}. 
\citet{2002A&A...392..991H} concluded that this dust must be located in 
the outer region of the gas disc, and follow-up studies using long baseline 
optical interferometry in the $H$ and $K$ spectral bands confirmed that the hot 
dust is confined within a skewed, asymmetric ring 
\citep{2009MNRAS.398.1309T}.

Based on the spectral properties and the existence of a molecular and dusty
disc, the primary of CI~Cam was suggested by \cite{1999A&A...348..888C} and \cite{2002A&A...392..991H}
 to belong to the group of B[e]SGs. Indication for the 
typical two-component B[e]SG wind \citep[see][]{1985A&A...143..421Z} was found 
by \citet{2002ApJ...565.1169R} and \citet{2002A&A...392..991H}. 
\citet{2008A&A...477..193M} discovered an extended faint H~$\alpha$ shell around 
CI~Cam. However, proper spectral classification of the B[e]SG component turned 
out to be difficult due to the uncertain distance and the absence of 
photospheric lines. \citet{2006ARep...50..664B} derived a spectral type of 
B4\,III-V, but the spectral energy distribution is also in agreement with 
an earlier classification of B0--B1\,III by \cite{2009MNRAS.398.1309T},
while \citet{2002A&A...392..991H} obtained B0--B2\,I and a luminosity of 
$\log L/$L$_{\odot} \ga 5.4$.

Our spectra for CI~Cam display numerous, narrow emission features, most of 
them single-peaked. The Paschen lines and the adjacent lines from the 
Ca\,{\sc ii} IR triplet appear slightly double-peaked. None of the 
lines in the spectral regions covered by our observations show the 
triple-peaked profiles reported by \citep{2002A&A...390..627M}, but it 
can also be caused by the lower resolution of our spectra.
The similarity of our data to those obtained in 1987 (the pre-outburst phase) by \citet{2003A&A...408..257Z}, who 
observed the same profile of the H~$\alpha$ line and reported on the lack of noticeable
[O\,{\sc i}] emission, indicates that the system is back in its pre-outburst phase.
This could mean 
that during the last few years, the conditions in the circumstellar matter have 
changed, i.e., parts of the ejected gaseous material probably expanded and 
diluted. Evidence for such a scenario comes also from recent observations in 
the near-infrared spectral range, which revealed that CI~Cam lost its CO band 
emission \citep{2014MNRAS.443..947L}. Expansion and dilution of the previously 
ejected material would also explain why we could not 
detect any emission in either [Ca\,{\sc ii}] or [O\,{\sc i}].


\subsection*{V1972~Cyg = MWC~342}

This star was identified by \citet{1933ApJ....78...87M} as an emission-line 
object, being either in the pre-main sequence \citep[Herbig Ae/Be 
star,][]{1990Afz....32..203B} or in an evolved evolutionary phase, 
perhaps as part of a binary system \citep{1999A&A...349..126M}.

Optical and infrared polarization is very small, hinting towards some slight asymmetries of 
the envelope \citep*{1990Afz....32..203B, 2005MNRAS.364..725O}, but  
providing no clear indication for presence of an ionized circumstellar disc.
\citet{2008A&A...477..193M} found a faint H~$\alpha$ shell around that star. 
$K$-band observations display strong emission of 
the hydrogen Pfund series, but no CO band emission (M. L. Arias, private 
communication). These characteristics are similar to the LBV candidate 
V1429~Aql \citep{2014MNRAS.443..947L} and other LBV stars surveyed 
in the near-infrared so far \citep{2013A&A...558A..17O}.

Our spectra of V1972~Cyg display strong emission of H~$\alpha$ and the Paschen 
lines. H~$\alpha$ is double-peaked with the red peak about 3.5 times more 
intense than the blue one. The Paschen lines display prominent broad, but 
single-peaked profiles with extended wings, which are strongly blended with the 
lines of the Ca\,{\sc ii} IR triplet. No [Ca\,{\sc ii}] lines
were detected. The [O\,{\sc i}] lines appear as intense, narrow, 
single-peaked emission features. Their possible split of 10\,km\,s$^{-1}$ 
reported by \citet{2003A&A...408..257Z} from high-resolution data, cannot be
resolved. Such a split of the forbidden emission lines implies that they are 
formed either in a rotating ring or an equatorial outflow of dense, neutral 
gas. The absence of the [Ca\,{\sc ii}] lines indicates that the circumstellar
gas must be less dense than in typical B[e]SGs, and the lack of molecular 
emission implies that the circumstellar material is not confined within a disc
or ring. 


\subsection*{V1429~Aql = MWC~314}

The object V1429~Aql was discovered by \citet{1927ApJ....65..286M} as a star 
showing bright iron lines. The stellar parameter determination via the Balmer 
discontinuum method failed due to the crowded emission line spectrum 
\citep*{2001A&A...368..160C}. Alternative methods revealed a temperature 
ranging from 32\,000\,K, based on the He\,{\sc i} lines, to 26\,700\,K, when 
considering the visible line spectrum and energy distribution. The latter is 
close to the value of 25\,000\,K suggested by \citet{1998A&AS..131..469M}, 
while \citet{2010A&A...517A..67C} found a much lower value of 16\,200\,K.

Earlier optical spectroscopic observations reported symmetric Balmer lines, 
He\,{\sc i} lines, and numerous, mainly double-peaked lines from permitted and 
forbidden transitions in neutral and low-ionized metals, including the lines of 
the Ca\,{\sc ii} IR triplet \citep{1998A&AS..131..469M, 
2010A&A...517A..67C}. The presence of the [Ca\,{\sc ii}] lines was suggested by 
\citet{1998A&AS..131..469M}, but their spectra suffer from telluric pollution, 
altering the shapes and strengths of these lines.

The double-peaked line profiles imply a non-spherical symmetry of the wind, and 
H~$\alpha$ imaging by \citet{2008A&A...477..193M} revealed a very large bipolar 
nebula structure with approximately 15\arcmin~in total length. Considering a 
distance of $\sim 3$\,kpc, this 
results in a total extension of the bipolar structure of 13.5\,pc from end to 
end, which is about 5 times larger than sizes of typical LBV nebulae. 
Bipolarity is also often observed in LBVs, and the morphology of the bipolar 
nebula around V1429~Aql strongly resembles the one seen around $\eta$~Car,
suggesting that a circumstellar disc could have caused the shaping of the 
nebula. A rotating disc hypothesis is also supported by the line formation studies by 
\citet*{2008A&A...487..637M}. However, if present at all, dust does not provide 
a major constituent of this disc because the spectral energy distribution shows 
no indication for an infrared excess due to circumstellar dust. The absence of 
measurable amount of dust implies that V1429~Aql is not a classical B[e]SG 
candidate.

Indeed, \citet{2013A&A...559A..16L} found that V1429~Aql is a massive 
semi-detached binary ($P=$~60.8\,d, eccentricity $e=$~0.23, total mass 
$M \simeq$~66\,$\pm$\,9\,M$_{\odot}$, and maximum distance between stars 
$a =$~1.22\,AU), consisting of a hot (18\,000\,K) and luminous (7.1$\times$ 
10$^{5}$\,L$_{\odot}$) primary star (possibly in an LBV-like state) with mass 
$\sim$\,40\,M$_{\odot}$, and a cool ($\sim$\,6300\,K) secondary with mass 
$\sim$\,26\,M$_{\odot}$. With an inclination angle of 17$\degr\pm$\,13$\degr$, the system 
is seen almost equator on, as is also inferred from the light curve characteristic for
an eclipsing system. The inclination angle also agrees with the 
value of 25$\degr\pm$\,5$\degr$ suggested by \citet{2008A&A...487..637M} for the 
orientation of the rotating (circumbinary) gas disc. 
The period of the system was confirmed by \citet{2015arXiv151006158F} who obtained $P=$~60.7\,d based on 
detailed analysis of absorption and emission lines in spectra covering a period of about 15 years.

Investigations of the near-infrared appearance of V1429~Aql revealed that
it shows strong emission in the hydrogen Pfund series, while no indication for
CO band emission could be found \citep{2014MNRAS.443..947L}. Such a behavior
seems to be typical for LBVs \citep{2013A&A...558A..17O}, supporting 
the LBV classification of V1429~Aql.

Both the LBV status of the primary as well as the masses of the binary components were recently questioned by \cite{2016MNRAS.455..244R}, based on combined spectroscopic, interferometric and photometric observations. While the results of \cite{2013A&A...559A..16L} are a possible solution, \cite{2016MNRAS.455..244R} propose a much lower total mass of the system of 20\,M$_{\sun}$ only and a mass ratio $M_{2}/M_{1} = 3$, and suggest that strong mass transfer in an interacting binary could have caused the reversal. However, the authors state that their investigations provide no unique solution and more observations are definitely needed to resolve this issue.

Our data show no clear evidence for [O\,{\sc i}] line emission in agreement 
with the earlier observations. This is another indicator that V1429~Aql is not a 
typical B[e]SG. H~$\alpha$ and the Paschen lines are in emission and have a 
symmetric single-peaked profile shape. The strength of the H~$\alpha$ line is 
comparable to the one of \citet{1998A&AS..131..469M}, but stronger than the one 
reported by \citet{2010A&A...517A..67C}. Variability in H~$\alpha$ and $R$-band 
polarization on time-scales of days to weeks has been notified by 
\citet{2006PASP..118..820W}. The origin of this variability could be due to 
the binarity \citep{2006PASP..118..820W, 2008A&A...487..637M, 
2012ASPC..465..358L}, while other studies suggest V1429~Aql might perform 
non-periodic pulsations as they appear in strange modes, typical 
for LBVs \citep{2011IAUS..272..422R}. Furthermore, our spectra display strong 
emission in both the Ca\,{\sc ii} IR triplet lines and the [Ca\,{\sc ii}] 
lines. While the profiles of the triplet lines appear symmetric and 
single-peaked, those of the forbidden lines are clearly double-peaked with the 
blue peak slightly more intense due to blending with adjacent emission lines.


\subsection*{OY~Gem = MWC~162 = HD~51585}

The object OY~Gem was discovered as an emission-line star by \citet{1933ApJ....78...87M}. Many forbidden 
emission lines (including lines of [O\,{\sc i}] and [Ca\,{\sc ii}]) from 
transitions covering a large range in excitation energies have been identified 
in extensive spectroscopic observations 
\citep*[see, e.g.,][]{1977A&A....56..143K, 1992SvAL...18..418A, 
1996A&AS..117..281J}. The object displays infrared excess emission 
\citep{1976A&A....47..293A}, but 
no indication for extended H~$\alpha$ emission  \citep{2008A&A...477..193M}.

Photometry and emission line strengths display strong variability on 
different time-scales \citep[see, e.g.,][]{2006AstL...32..594A}, which are 
interpreted as due to variability in the stellar wind. 
\citet*{2010MNRAS.401.1199W} found indication for a companion based on 
spectroastrometric observations. They list OY~Gem as a Herbig Ae/Be star. 
However, Herbig Ae/Be and post-AGB stars have many common emission features in their 
spectra. And in fact, high-ionization forbidden emission lines, the presence of two 
detached shells \citep{1992SvAL...18..418A, 2008MNRAS.385.1076V} with hot 
(1100--1200\,K) and cold (150--180\,K) carbon-rich dust detected with Spitzer 
\citep{2009ApJ...703..585C}, and the absence of any spectroscopic evidence 
for accretion over the last decades clearly speak in favor of classification as 
post-AGB or compact planetary nebula rather than as a pre-main sequence object.

The forbidden lines in our spectra of OY~Gem are sharp and single peaked. The 
exception is H~$\alpha$, which has a double-peaked profile with the red peak 
$\sim$\,3.5 times the intensity of the blue one. The intensity of the H~$\alpha$ 
line in OY~Gem is among the strongest in our sample. Furthermore, the wings of 
the H~$\alpha$ line extend to large velocities ($\sim$\,3200\,$\pm$\,100\,km\,s$^{-1}$). Such broad wings (of $\sim$\,2900\,km\,s$^{-1}$) have also 
been reported by \citet{1981A&A...103L...3S}, not only for H~$\alpha$, but also 
for higher Balmer lines. In contrast, the Paschen lines are very narrow and 
only slightly blended with the equally narrow Ca\,{\sc ii} IR triplet 
lines. OY~Gem has by far the strongest [O\,{\sc i}] and [Ca\,{\sc ii}] lines 
of all our sample stars, comparable to those typically seen in post-AGB stars.

\begin{figure*}
\includegraphics{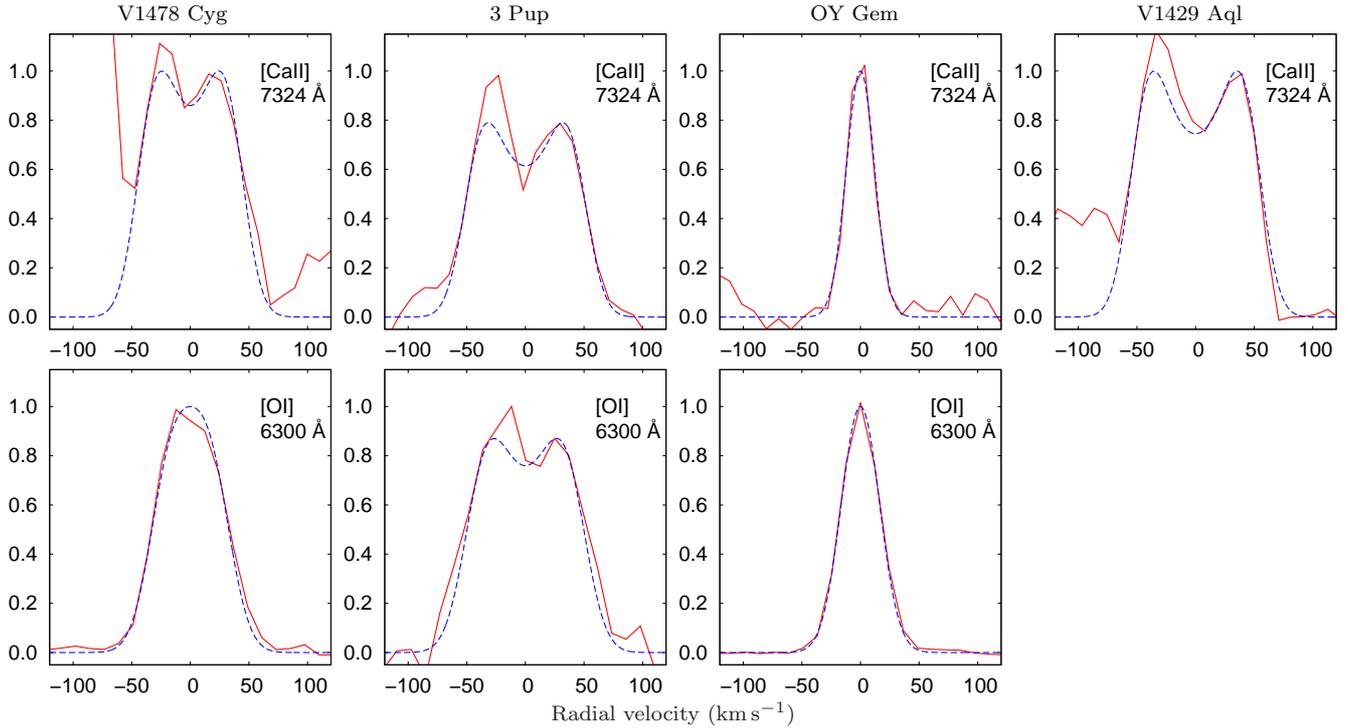}\\
Radial velocity (km\,s$^{-1}$)
\caption{Kinematic model fits (dashed blue) to the observed profiles (solid red) of the forbidden lines.}
\label{fits}
\end{figure*}


\subsection*{V743~Mon = MWC~158 = HD~50138}

The star V743~Mon belonged to the group of unclassified B[e] 
stars for a long time. It displays strong spectral as well as small-scale photometric 
variabilities, which are suggested to be caused by shell phases of the star 
\citep[e.g.,][]{1991IAUC.5164....3A, 2009A&A...508..309B}, and at least one
larger outburst was recorded in 1978/79 \citep{1985A&AS...60..373H}.
However, no extended H~$\alpha$ emission was detected by
\citet{2008A&A...477..193M}.

\begin{table}
\caption{Disc kinematics.}
\label{tab:disc}
\begin{tabular}{l @{~~}c @{~~}l @{~~}c @{~~}l}
\hline
             &\multicolumn{2}{c}{[Ca\,{\sc ii}]}            &\multicolumn{2}{c}{[O\,{\sc i}]}                        \\
Star         &$v_{\rm rot}$[km\,s$^{-1}$]&~~R [AU]            &$v_{\rm rot}$[km\,s$^{-1}$] &~~R [AU]  \\
\hline
V1478 Cyg    &38$\pm$1                   &24.6$\pm$1.3      &25$\pm$1                    &55.7$\pm$5.9        \\
3~Pup        &72$\pm$1                   &\phantom{2}3.0$\pm$0.4$^a$ &68$\pm$1               &\phantom{5}3.4$\pm$0.5$^a$       \\
               &                        &1.67$\pm$0.13$^b$ &                   &1.87$\pm$0.14$^b$       \\
V1429 Aql    &50$\pm$1                   &23.4$\pm$4.0      &$-$                         &\phantom{55.7}$-$      \\
OY Gem       &0                          & \phantom{23.4}$-$                 &0                           &\phantom{55.7}$-$        \\
\hline
\end{tabular}

$^a$ Using the mass range of \citet{2011A&A...526A.107M};\\
$^b$ using a mass range of 9--10.5 M$_{\odot}$, see Sect. \ref{sec:disc}.
\end{table}

V743~Mon has often been referred to as a Herbig (hence pre-main-sequence) star, but 
the lack of a close-by star-forming region in combination with regular shell 
phases of the star, which are not observed in pre-main-sequence stars, make 
such a classification rather doubtful.  
V743~Mon was recently re-classified by 
\citet{2009A&A...508..309B} as a B6--7\,III--V star of $\sim$\,5\,M$_{\odot}$, 
located at the end of the main sequence. An extended, dusty 
disc, seen under an inclination of 56$\degr\pm$\,4$\degr$, was discovered by 
\citet{2011A&A...528A..20B}. Based on spectroastrometric observations, 
\citet{2006MNRAS.367..737B} suggested that V743~Mon could be a binary system. 
However, no indication for a companion has been found from either spectroscopic 
or interferometric observations \citep{2011A&A...528A..20B}. As an 
additional curiosity, the recent analysis of line-profile variability of 
photospheric lines suggests that the star could be pulsating 
\citep{2012A&A...548A..13B}. The confirmation is, however, still pending.

The shape of the H~$\alpha$ line profile agrees with the presence of a gaseous 
disc seen at intermediate inclination. The Paschen lines are broad and seem to 
consist of multiple components. Emission in the Ca\,{\sc ii} IR triplet 
lines seems to be present, but these lines are strongly blended with the Paschen lines, while
the [Ca\,{\sc ii}] lines are absent. The [O\,{\sc i}] lines are narrow and 
single-peaked.


\subsection*{BD+23~3183 = IRAS~17449+2320}

Not much is known yet about the object BD+23~3183. It was discovered as a new 
H~$\alpha$ emission-line star by \citet{1986ApJ...300..779S}, who listed it as 
number 145 in his catalogue. \citet{1988AJ.....96..777D} performed a follow-up 
low-resolution optical spectroscopic study in which they classify it as Be 
star. The detection of a strong infrared excess and of [O\,{\sc i}] emission 
resulted in the classification as B[e] star \citep{2007ApJ...671..828M}. 
Other lines such as Fe\,{\sc ii}, He\,{\sc i}, and Mg\,{\sc ii} were in pure 
absorption, suggesting a spectral type of A0 and classification as a dwarf 
star. 

Our spectra confirm the presence of [O\,{\sc i}] and Balmer line emission. 
However, compared to the observations by \citet{2007ApJ...671..828M} the 
blue peak of our H~$\alpha$ line is much less intense. The Paschen lines display 
small and narrow central emission features superimposed on the broad 
photospheric absorption component similarly to what was seen for H~$\beta$ 
\citep{2007ApJ...671..828M}. The H~$\alpha$ emission line shows also 
indication for an underlying photospheric absorption component, extending to 
higher velocities than the emission wings. The spectra show no evidence for 
calcium line emission, neither in [Ca\,{\sc ii}] nor in the Ca\,{\sc ii} 
IR triplet.


\subsection*{HD~281192 = MWC~728}

The object HD~281192 was first detected as an emission-line star by \citet{1949ApJ...110..387M} and listed in their 
second supplement to the Mount Wilson Catalogue of early-type emission-line 
stars. Although HD~281192 shares its IRAS colours with typical OH/IR stars, no 
OH maser lines were detected by  \citet{1993ApJS...89..189C}. 
\citet{2007ApJ...671..828M} obtained high-resolution optical spectra and 
reported the emission of double-peaked profiles of [O\,{\sc i}] and the Balmer 
lines, while the iron lines were in pure absorption. 
Furthermore, from the presence of weak absorption 
features of Li\,{\sc i} and Ca\,{\sc i} these authors propose a binary nature 
of HD~281192. Support for a companion is also provided by recent $K$-band 
spectra, which revealed CO bands in absorption (Arias et al., in preparation), 
typically seen from the atmospheres of late-type giants. 
Indeed, \cite{2015ApJ...809..129M} found that the system consists of
a B5\,Ve primary and a G8\,III secondary, and has an orbital period of 27.5\,d.
They also report on non-periodic variability of the emission lines.

Our spectra confirm the presence of pronounced, narrow [O\,{\sc i}] lines and a 
double-peaked H~$\alpha$ emission with the red peak about twice as intense as 
the blue peak. The emission is superimposed on a much wider absorption. Our 
[O\,{\sc i}] lines are single-peaked, but our spectra have lower resolution 
than those obtained by \citet{2007ApJ...671..828M,2015ApJ...809..129M}. The shape of the 
emission features in our spectra are very similar to those in BD+23~3183, but 
the intensities of H~$\alpha$ and the [O\,{\sc i}] lines in BD+23~3183 are 
approximately twice and half as strong, respectively. No indication for
[Ca\,{\sc ii}] or the Ca\,{\sc ii} IR triplet is found.


\section{Discussion}
\label{sec:disc}

The former studies of forbidden emission lines in a 
sample of B[e]SGs by \citet{2012MNRAS.423..284A}  
revealed that the kinematics, derived from the line profiles, 
agrees with the scenario of a Keplerian rotating disc, in which the [Ca\,{\sc ii}]
lines are formed closest to the star, followed by the [O\,{\sc i}] $\lambda$5577 line
and the [O\,{\sc i}] $\lambda\lambda$6300, 6364 lines. The hypothesis of Keplerian rotation
of these circumstellar discs is further supported by near-infrared studies,
focused on the molecular CO band emission, which shows indication for 
rotational broadening as well \citep{2012A&A...548A..72C,2013A&A...558A..17O,
2012A&A...543A..77W, 2013A&A...549A..28K, 2012ASPC..464...67M,
2015AJ....149...13M}. 

Only three objects in our sample display emission in the two strategic sets of
forbidden lines, [O\,{\sc i}] and [Ca\,{\sc ii}]. These are the two B[e]SGs 
V1478~Cyg and 3~Pup, and the compact planetary nebula star OY~Gem. The LBV 
candidate V1429~Aql only displays the [Ca\,{\sc ii}] lines, but no [O\,{\sc i}]
emission. Except for OY~Gem, all these stars are confirmed disc sources (see 
Table\,\ref{tab:stel}), and the line profiles clearly show kinematic broadening.
The remaining object with confirmed disc in our sample, V743~Mon, has only 
the B[e] star characteristic [O\,{\sc i}] emission. As the line profile is single-peaked, 
no reliable constraints with respect to the kinematics in the line-forming region can be obtained.

In the remaining, we focus on the objects with [Ca\,{\sc ii}] emission.
For clearer inspection of the line profile shapes, we subtracted the continuum 
and scaled the line intensities to unity. These profiles are displayed in 
Fig.\,\ref{fits}. In both B[e]SGs, the [Ca\,{\sc ii}] line is clearly
double-peaked and broader than the [O\,{\sc i}] line, in agreement with
the Keplerian disc scenario for this type of objects. Also in V1429~Aql, the
[Ca\,{\sc ii}] lines have double-peaked profiles. Because our spectra
have only medium resolution ($\sim$\,23\,km\,s$^{-1}$ at [O\,{\sc i}] and 
$\sim$\,20\,km\,s$^{-1}$ at [Ca\,{\sc ii}]), the peak separation, listed in 
Table\,\ref{tab:forbidden}, does not provide proper rotational velocities. 
Therefore, we applied a simple, purely kinematic model to compute line profiles 
for comparison with the observations. Assuming that the emission originates 
from a narrow (i.e. constant velocity) rotating ring, we calculate the profile 
shape considering only rotational velocity, projected to the line of sight 
according to the observed inclination angles (Table\,\ref{tab:stel}), and the 
resolution of the spectrograph. In all cases, we obtain good fits to the 
observed line profiles, demonstrating that the observed emission originates 
indeed from a narrow ring region. These fits are included in Fig.\,\ref{fits}, 
and the corresponding rotational velocities are listed in Table\,\ref{tab:disc}.
Also listed are the corresponding radii of the emitting rings assuming Keplerian rotation.

For OY~Gem, no kinematic broadening beyond the spectral resolution was needed
to fit the line profile shapes. If this object has a disc, it must be oriented
pole-on. Or, if the emission originates from a shell, that was ejected during 
the former AGB phase, the expansion velocity of the material must have 
decelerated to (almost) zero. High-resolution spectra are clearly needed 
to properly resolve profile shapes and to investigate the kinematics.  

Despite the reported disc around the LBV candidate V1429~Aql (MWC~314) and
the detection of the [Ca\,{\sc ii}] lines, no [O\,{\sc i}] lines are observed. This 
could indicate that the gas disc around this massive, luminous object is too 
hot for hydrogen (hence oxygen) to exist in its neutral state at an amount 
sufficient for detection. Considering the mass estimates and binary separation by
\citet{2013A&A...559A..16L}, the emission region of [Ca\,{\sc ii}] must clearly 
be circumbinary, and from our rotational velocity we obtain a distance of 
23.4\,$\pm$\,4\,AU.

The B[e]SG V1478~Cyg (MWC~349A) is surrounded by a large (0.05\,AU to 
130\,AU) Keplerian rotating disc, consisting of an ionized surface layer and a 
cool, neutral and dusty mid-layer, at least in the outer regions.
Moreover, a wind is ejected from the ionized disc at a radius of $\sim$\,24\,AU, with a terminal 
outflow velocity of 60\,km\,s$^{-1}$, and rotating in the same sense 
as the disc \citep{2013A&A...553A..45B, 2014A&A...571L...4B}. With the
recent, new mass estimates of 38--40\,M$_{\odot}$ \citep{2012A&A...541A...7G, 
2013A&A...553A..45B}, we locate the [Ca\,{\sc ii}] emission region at a        
distance of 24.6\,$\pm$\,1.3\,AU, which coincides with (or is just beyond) the 
region, from which the wind is launched. The Keplerian velocity obtained for
[O\,{\sc i}] agrees with the values obtained for several hydrogen recombination 
maser lines \citep[e.g.,][]{1994A&A...283..582T, 1994A&A...288L..25T, 
1995A&A...300..843T}. The distance of their emission regions amounts to
55.7\,$\pm$\,5.9\,AU. 

The rotational velocities of the forbidden lines in 3~Pup, when combined with 
those obtained from CO band \citep[53\,km\,s$^{-1}$,][]{2012ASPC..464...67M}
and SiO band emission \citep[48\,km\,s$^{-1}$,][]{2015ApJ...800L..20K},
provide the best example for tracing the Keplerian gas disc from the atomic to
the molecular region. However, proper distance determination of the emission
regions is hampered by the uncertain mass (luminosity) of the central object. 
\citet{2011A&A...526A.107M} determined a value of 15--20 M$_{\odot}$ and a distance of 650\,pc.
The kinematically resolved gas disc would thus extend from 
3.0\,$\pm$\,0.4\,AU ([Ca\,{\sc ii}]) to 6.7\,$\pm$\,1.0\,AU (SiO). 
This parameter combination would place (parts of) the atomic and 
molecular emission regions beyond the inner edge of the dusty disc, which is 
suggested to be at about 4\,AU \citep{2011A&A...526A.107M}. However, the
temperatures of these molecular regions clearly exceed the dust sublimation 
temperature, and should hence be located closer to the star than the dusty
part. 

This inconsistency can only be resolved for either a lower mass of the central object or a greater distance (or a combination of both). \cite{2010AstBu..65..150C} have determined a slightly greater distance of 700\,pc, a spectral type of A2.7$\pm$0.3\,Ib, and a value of $M_{V}=-$5.5\,$\pm$\,0.3. This results in a luminosity value of $\log L/$L$_{\odot}=$~4.10\,$\pm$\,0.12. Assigning the total luminosity to the massive component, a mass can be estimated from a comparison with stellar evolutionary tracks from \cite{2012A&A...537A.146E} for single rotating stars with solar metallicity. This delivers a possible mass range of 9--10.5\,M$_{\odot}$. The kinematically traced disc would thus extend from 1.67\,$\pm$\,0.13\,AU ([Ca\,{\sc ii}]) to 3.75\,$\pm$\,0.29\,AU (SiO), while the slightly greater distance to the object would place the inner edge of the dusty disc at 4.2\,AU, hence providing a more consistent scenario.

                                                                                
\section{Conclusions}
\label{sec:concl}

In our survey of B[e] stars, we detect emission of the strategic [Ca\,{\sc ii}]
lines in four out of a sample of nine objects. Three of these objects with [Ca\,{\sc ii}] emission are known to
possess circumstellar discs, and the kinematics derived from the line profiles
agree with Keplerian rotation. Two of these sources, V1478~Cyg and 3~Pup, are 
B[e]SGs, while the third one, V1429~Aql, is an LBV candidate. The fourth object, 
OY~Gem, is a known compact planetary nebula. If it has a disc, it should be 
oriented pole-on, because no kinematic broadening was detected beyond spectral 
resolution. 

Another object of this survey is CI~Cam, which was reported 
by \cite{1999A&A...348..888C} and \cite{2002A&A...392..991H} to show all characteristics typically seen in B[e]SGs
after it had experienced an outburst. This star shows no indication for either 
[Ca\,{\sc ii}] or [O\,{\sc i}] emission in our spectra, in line with the 
reported disappearance of its CO band emission. This suggests that CI~Cam
is back in its pre-outburst state.

Our sample also contains three B[e] stars with unclear evolutionary
stage. None of them has the [Ca\,{\sc ii}] lines, and only the disc source 
V743~Mon displays emission in the Ca\,{\sc ii} IR triplet. 
Based on the results of \citet{2012MNRAS.423..284A} and also those presented here, the [CaII] lines tend to appear in regions closer to the star than the [OI] lines, i.e., in regions of higher density. The lack of measurable amounts of emission from these non-massive stars might hence indicate that their circumstellar environments have not the proper physical conditions in terms of density and size of the emitting volume.

The detection of the strategic forbidden lines, particularly in massive stars 
which have circumstellar environments with high density and rather large volume, as it is the case in the discs of B[e]SGs, strengthens their importance as valuable disc tracers. 
Hence, these lines provide an ideal tool to identify 
objects with similar 
circumstellar environments, and allow us to study the physical conditions and 
kinematics within their line forming regions.


\section*{Acknowledgments}

We thank the technical  staff at the Ond{\v r}ejov  Observatory
for  the  support  during  the  observations.
We thank the referee for useful comments that helped to improve the presentation of our results.
This research made use of the NASA Astrophysics Data System (ADS)
and of the SIMBAD database operated at CDS, Strasbourg, France.
M.K. acknowledges financial support from GA\,\v{C}R under grant number
14-21373S and from the European Structural Funds grant for the Centre of
Excellence "Dark Matter in (Astro)particle Physics and Cosmology".
The Astronomical Institute Ond\v{r}ejov is supported by
the project RVO:67985815. 
A.A. acknowledges financial support from Estonian Science Foundation grant ETF8906.
This work was also supported by the research
projects SF0060030s08 and IUT40-1 of the Estonian Ministry of Education and Research.

\bsp
\label{lastpage}

\end{document}